\documentclass{elsart}

\usepackage{psfig}

\usepackage{natbib}

\begin{document}

\runauthor{Murayama, Nagao, and Taniguchi}


\begin{frontmatter}

\title{How Do We See the Nuclear Region (r $<$ 0.1 pc) of
       Narrow-Line Seyfert 1 Galaxies?}

\author{Takashi Murayama\thanksref{Murayama}},
\author{Tohru Nagao\thanksref{Nagao}},
\author{Yoshiaki Taniguchi\thanksref{Taniguchi}}
\address{Astronomical Institute, Graduate School of Science,
Tohoku University, Sendai 980-8578, Japan}
\thanks[Murayama]{E-mail: murayama@astr.tohoku.ac.jp}
\thanks[Nagao]{E-mail: tohru@astr.tohoku.ac.jp}
\thanks[Taniguchi]{E-mail: tani@astr.tohoku.ac.jp}

\begin{abstract}
We propose two statistical tests to investigate
how we see the nuclear region ($r < 0.1$ pc) of narrow-line
Seyfert 1 galaxies (NLS1s).
1) The high-ionization nuclear emission-line
region (HINER) test: Seyfert 1 galaxies (S1s) have systematically higher flux
ratios of [Fe\thinspace\textsc{vii}] $\lambda$6087 to
[O\thinspace\textsc{iii}] $\lambda$5007 than Seyfert 2 galaxies (S2s).
This is interpreted in that a significant part of
the [Fe\thinspace\textsc{vii}] $\lambda$6087 emission arises
from the  inner
walls of dusty tori that  cannot be seen in S2s
(Murayama \& Taniguchi 1998a,b).
2) The mid-infrared test: S1s have systematically
higher flux ratios of the $L$ band (3.5 $\mu$m) to the \textit{IRAS} 25
$\mu$m band than S2s. This is also interpreted in
that a significant part of the $L$ band emission arises
from the inner walls of dusty tori, because the tori
are optically thick enough to absorb the $L$ band
emission if the tori are viewed  nearly edge on
(Murayama et al.\ 2000).
Applying these tests to a sample of NLS1s, we have
found that the NLS1s possibly have nearly the same properties as
S1s.
\end{abstract}

\end{frontmatter}


\section{Introduction}
Dusty tori around active galactic nuclei (AGNs) play an important role
in the classification of Seyfert galaxies
(Antonucci \& Miller 1985; see also Antonucci 1993 for a review).
Seyfert galaxies observed  face-on to the torus are
recognized as Seyfert 1 galaxies (S1s) while those observed edge-on
are recognized as Seyfert 2 galaxies (S2s).
Therefore, physical properties of dusty tori are of great interest.
We introduce two statistical studies searching for 
any systematic differences of viewing angles toward dusty tori
among narrow-line Seyfert 1 galaxies (NLS1s), S1s, and S2s:
1) ionization condition of the inner wall of the torus based on
   high-ionization emission lines
   (Murayama \& Taniguchi 1998a,b),
and
2) viewing angle towards the dusty torus based on mid-infrared color
   (Murayama et al.\ 2000).
These tests may provide constraints 
for models of NLS1s (for a review, see Boller et al.\ 1996),
especially the viewing-angle-dependent unified model.

\section{High-Ionization Emission Lines in Seyfert Galaxies}
Optical spectra of AGNs often show 
very high-ionization emission lines such as [Fe\thinspace\textsc{vii}],
[Fe\thinspace\textsc{x}], and [Fe\thinspace\textsc{xiv}]
(the so-called coronal lines).
Since the inner wall of the torus is exposed to intense radiation
from the central engine, it is naturally expected that the wall 
can be one of the important parts of  the high-ionization
nuclear emission-line region (HINER) (Pier \& Voit 1995).
If the inner wall is an important part  of the HINER, 
it should be expected that S1s would tend to have 
more intense HINER emission, because in S2s the inner wall would be 
obscured by the torus itself.
Actually, Murayama \& Taniguchi (1998a) showed that 
S1s have systematically higher [Fe\thinspace\textsc{vii}]
$\lambda$6087/[O\thinspace\textsc{iii}] $\lambda$5007 intensity ratios
than S2s, while the  [O\thinspace\textsc{iii}] $\lambda$5007 luminosity is
almost the same in these two Seyfert types.
Therefore, the excess [Fe\thinspace\textsc{vii}] emission
in S1s may arise from the inner wall of the torus
that cannot be seen in S2s, because S2s have tori that are viewed edge-on.

In order to investigate whether or not NLS1s have such excess emission
of [Fe\thinspace\textsc{vii}],
we compiled emission-line spectral data for 11
NLS1s from the literature (see Nagao et al.\ 2000)
and compared 
[Fe\thinspace\textsc{vii}]/[O\thinspace\textsc{iii}] ratios
with those of 21 S1s and 34 S2s (Fig.~1).
[Fe\thinspace\textsc{vii}]/[O\thinspace\textsc{iii}] ratios of the NLS1s are
systematically higher than those of the S2s, but appear to be a little lower
than those of the S1s.
If we apply the Kolmogrov-Smirnov (KS) test,
the probability that the observed distributions
of the NLS1s and the S2s originate from the same underlying population
is 0.011\%, while that for the NLS1s and the S1s is 4.7\%.
Though the difference of the distribution between the NLS1s and the S1s
is unclear statistically, 
we may conclude that NLS1s have an excess of [Fe\thinspace\textsc{vii}] emission,
 like the S1s.

\begin{figure}[tb]
\centerline{\psfig{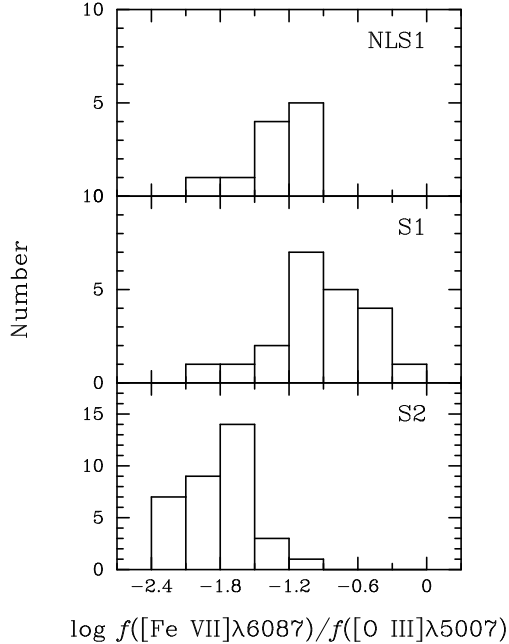}}
\caption{Frequency distributions of the
[Fe\thinspace\textsc{vii}]$\lambda$6087/[O\thinspace\textsc{iii}]$\lambda$5007
intensity ratios among the NLS1s, the S1s and the S2s.
}
\end{figure}

\section{The Mid-Infrared Diagnostics}
Because of the anisotropic properties of
the dusty torus emission (e.g., Pier \& Krolik 1992, 1993),
the emission at $\lambda <$ 10 $\mu$m is systematically stronger
in type 1 AGNs than in type 2s, while emission at $\lambda >$ 20 $\mu$m
is not significantly different between type 1 and type 2 AGNs.
Recently, Murayama et al.\ (2000) showed that 
the flux ratio of the $L$ band to the \textit{IRAS} 25 $\mu$m
band (\textit{L/IRAS}25) can act as an estimate of the viewing angle
toward the torus.
S1s have systematically higher \textit{L/IRAS}25 ratios than S2s
because the inner hotter region can been seen in the S1s.
Therefore, we investigate \textit{L/IRAS}25 ratios for NLS1s
to determine whether NLS1s have face-on tori like S1s
or  edge-on tori like S2s.
We compiled photometric data of $L$ and \textit{IRAS} 25 $\mu$m
for 9 NLS1s, 24 S1s and 5 S2s. The results are shown in Fig.~2.
The distribution of \textit{L/IRAS}25 ratios of the NLS1s
seems more like that of the S1s than that of the S2s,
although the results are not statistically significant
because the sample is not large enough.
The KS probability that the observed distributions
of the NLS1s and the S2s originate in the same underlying population
is 4.9\% while that for the NLS1s and the S1s is 61.7\%.

\begin{figure}[tb]
\centerline{\psfig{figure=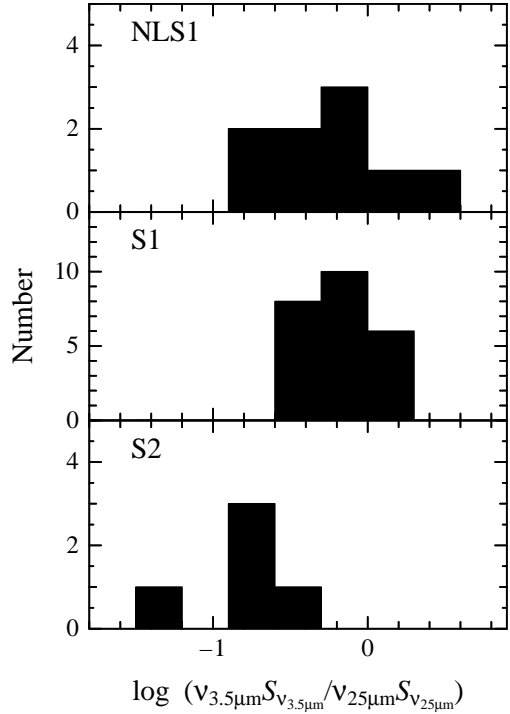,width=2.6truein}}
\caption{Frequency distributions of the flux ratio
of 3.5 $\mu$m to 25 $\mu$m for the NLS1s, the S1s, and the S2s.
}
\end{figure}

\section{Discussion}
Although the mid-infrared diagnostics give only marginal results,
both  tests show that the NLS1s tend to have properties
unlike the S2s.
Therefore, we may conclude at least that the NLS1s do not have
edge-on tori.
It is not clear whether or not there is a difference
in viewing angle between the NLS1s and the S1s
because of the statistically small data sample.
Although, of course further observational data is necessary, 
the present results possibly show that the NLS1s have nearly the same
properties as the S1s and that the NLS1s might have almost the same
viewing angle distribution  as S1s.
This might support the model of a low-mass central black hole
with a high accretion rate for the NLS1s.
However, the viewing angle we examined here is that of the dusty torus,
not that of the broad-line region (BLR) disk or the accretion disk.
It is still possible that the BLR disk of the NLS1s is always observed
from a smaller viewing angle than that of the normal S1s,
if the BLR disk is randomly oriented with respect  to the torus.
  


\end{document}